\documentclass[usenatbib]{mn2e}

\usepackage{graphicx}

\title[GRB\,060206 and the quandary of achromatic breaks]
{GRB\,060206 and the quandary of achromatic breaks in afterglow light curves}

\author[P.A.~Curran et al.]
{P.A.~Curran$^1$\thanks{e-mail: pcurran@science.uva.nl}, 
A.J.~van~der~Horst$^1$,
R.A.M.J.~Wijers$^1$, 
R.L.C.~Starling$^2$, \newauthor
A.J.~Castro-Tirado$^3$,
J.P.U.~Fynbo$^4$, 
J.~Gorosabel$^3$,
A.S.~J\"arvinen$^{5,6}$,\newauthor
D.~Malesani$^{4,7}$,
E.~Rol$^2$,
N.R.~Tanvir$^2$,
K.~Wiersema$^1$,
M.R.~Burleigh$^2$,\newauthor
S.L.~Casewell$^2$,
P.D.~Dobbie$^8$,
S.~Guziy$^9$, 
P.~Jakobsson$^{10}$,
M.~Jel\'inek$^3$, \newauthor
P.~Laursen$^4$,
A.J.~Levan$^{11}$, 
C.G.~Mundell$^{12}$, 
J.~N\"ar\"anen$^{13}$,
S.~Piranomonte$^{14}$ \\
$^1$ Astronomical~Institute, University~of~Amsterdam, Kruislaan~403, 1098~SJ~Amsterdam, The~Netherlands\\
$^2$ Department~of~Physics~and~Astronomy, University~of~Leicester, University~Road, Leicester~LE1~7RH, UK\\
$^3$ Instituto~de~Astrof\'isica~de~Andaluc\'ia~(IAA-CSIC), PO~Box~3004, 18080~Granada,~Spain\\
$^4$ Dark~Cosmology~Centre, Niels~Bohr~Institute, University~of~Copenhagen, Juliane~Maries~Vej~30, 2100~Copenhagen, Denmark\\
$^5$ Astrophysikalisches Institut Potsdam, An der Sternwarte 16, D-14482  Potsdam, Germany \\
$^6$ Astronomy Division, PO Box 3000, FI-90014 University of Oulu, Finland \\
$^7$ International School for Advanced Studies (SISSA/ISAS), via Beirut 2-4, I-34014 Trieste, Italy \\
$^8$ Anglo-Australian Observatory, PO Box 296, Epping, NSW1710, Australia \\
$^9$ Kalinenkov Astronomical Observatory, Nikolaev State University, Nikolskaya 24, 54030, Ukraine \\
$^{10}$ Centre for Astrophysics Research, University of Hertfordshire, College Lane, Hatfield, Herts AL10 9AB, UK \\
$^{11}$ Department of Physics, University of Warwick, Coventry, CV4 7AL, UK\\
$^{12}$ Astrophysics Research Institute, Liverpool John Moores University, Twelve Quays House, Birkenhead, CH41 1LD, UK\\
$^{13}$ Nordic Optical Telescope, Apartado 474, Santa Cruz de La Palma, Spain \\
$^{14}$ INAF, Osservatorio Astronomico di Roma, via Frascati 33, I-00040, Monteporzio Catone (Roma), Italy \\
}

\begin{document}

\date{Accepted 2007 July 17. Received 2007 June 26; in original form 2007 June 08}

\pagerange{\pageref{firstpage}--\pageref{lastpage}} \pubyear{}

\maketitle

\label{firstpage}


\begin{abstract}
Gamma-ray burst afterglow observations in the \emph{Swift} era have
a perceived lack of achromatic jet breaks compared to the \emph{BeppoSAX} era. 
We present our multi-wavelength analysis of GRB\,060206 as an illustrative
example of how inferences of jet breaks from optical and X-ray data might differ.
The results of temporal and spectral analyses are compared, and attempts are made to fit the data
within the context of the standard blast wave model. We find that while 
the break appears more pronounced in the optical and evidence for it
from the X-ray alone is weak, the data are actually
consistent with an achromatic break at about 16 hours. This break and the
light curves fit standard blast wave models, either as a jet break or as an injection break. 
As the  pre-\emph{Swift} sample of afterglows are dominated by optical observations, 
and in the \emph{Swift} era most well sampled light curves are in the X-ray, 
caution is needed when making a direct comparison between the two samples, 
and when making definite statements on the absence of achromatic breaks. 
\end{abstract}



\section{Introduction}\label{section:introduction}

Gamma-Ray Bursts (GRBs) are well described by the blast wave, or fireball, model
\citep{rees1992:mnras258,meszaros1998:apj499}, which details their temporal
and spectral behaviour.  In this model GRB afterglow emission is created by
shocks when a collimated ultra-relativistic jet ploughs
into the circumburst medium, driving a blast wave ahead of it.
This causes a non-thermal spectrum widely
accepted to be synchrotron emission, with characteristic power-law slopes
and spectral break frequencies.  The signature of the collimation is an
achromatic temporal steepening or `jet break' at $\sim 1$ day in an
otherwise decaying, power-law light curve.  The level of collimation, or
jet opening angle, has important implications for the energetics of the
underlying physical process.

Since the launch of the \emph{Swift} satellite \citep{gehrels2004:ApJ611},
this standard picture has been called into question by 
the rich and novel phenomena discovered in the both the early and late light curves 
(e.g., \citealt{nousek2006:ApJ642}). 
Here we focus on the perceived lack of achromatic temporal breaks in the \emph{Swift} era,
up to weeks in some bursts (e.g., \citealt{panaitescu2006:MNRAS369,burrows2007:astro.ph2633}), 
which calls into question the effects of
collimation and therefore the energy requirements of progenitor models.
Some bursts show no evidence for breaks in either optical or X-ray, while
others show clear breaks in one regime without any apparent accompanying break in the other.
Even in those bursts where an achromatic break is observed, they may not be
consistent with a jet break as predicted by the blast wave model
(e.g., GRB\,060124, \citealt{curran2006:astro.ph.10067}).  
We should note that our expectations of the observable signature of a jet break, 
including the fact that it ought to be perfectly achromatic, is based on highly simplified
models, notably those of \cite{rhoads1997:ApJ487,rhoads1999:ApJ525} and \cite{sari1999:ApJ519},
and break observations, pre-\emph{Swift}, that were based predominately in one regime (i.e., optical).
So apart from well sampled multi-regime observations, more realistic models and simulations
of the light curves, beyond the scope of this Letter, will also be required to settle this issue.

As the apparent lack of observed achromatic breaks is an important issue in the \emph{Swift} era, 
we will discuss the perceived presence and absence of these achromatic breaks, using the
long burst GRB\,060206 as an illustrative example. 
We present our multi-wavelength analysis of the well sampled afterglow from X-ray to optical wavelengths.  
In \S\ref{section:observations} we introduce our observations while in \S\ref{section:results} we
present the results of our temporal and spectral analyses.  
In \S\ref{section:discussion} we discuss these results in the overall context
of the blast wave model of GRBs and we summarise our findings in \S\ref{section:conclusion}.


\section{Observations}\label{section:observations}

Throughout, we use the convention that a power-law flux is given as $F_{\nu} \propto t^{-\alpha} \nu^{-\beta}$
where $\alpha$ is the temporal decay index and $\beta$ is the spectral index. 
All errors and uncertainties are quoted at the $1\sigma$ confidence level.  

\subsection{Optical}\label{section:obs-opt}

Optical observations in $B$, $V$, $R$ and $I$ bands were obtained at the 2.5\,m Nordic Optical Telescope (NOT), 2.5\,m Isaac Newton Telescope (INT) and 3.6\,m Telescopio Nazionale Galileo (TNG) on La Palma, the 1.5\,m Observatorio de Sierra Nevada (OSN) in Granada, Spain,  the 1.8\,m Astrophysical Observatory of Asiago, Italy and the 2.0\,m Faulkes Telescope North (FTN) at Haleakala, Hawaii (Table \ref{optical_log}).
The optical counterpart was identified in initial $R$ band frames, however no counterpart was detected in the $B$ band frames, in agreement with the significant level of line blanketing associated with the Lyman forest at a redshift of $z = 4.048$ \citep{fynbo2006:aa451}: the fluxes of the  $B$, $V$ and $R$ bands are reduced to 8, 50 and 88 per cent, respectively, of their true values \citep{madau1995:ApJ441}. 
The field was calibrated via a standard \cite{landolt1992:AJ104} field taken by the OSN on a photometric night.
Differential photometry was carried out relative to a number of stars within $\sim 5^{\prime}$ of the burst, with resulting deviations less than the individual errors. The photometric calibration error is included in error estimates.
We combine our $R$ band data with that already published from the RAPTOR \& MDM telescopes (\citealt{wozniak2006:apj642,stanek2007:apj654}; where MDM was shifted $+0.22$ magnitudes as in \citealt{monfardini2006:apj648}) to extend the optical light curve past $1 \times 10^6$\,s since trigger.

\begin{table}	
  \centering	
  \caption{Optical observations of GRB\,060206. Magnitudes are given with $1\sigma$ errors or as $3\sigma$ limits.} 	
    \label{optical_log} 	
    \begin{tabular}{l l l l} 
      $T_{\mathrm{mid}}$  & $T_{\mathrm{exp}}$ & Band & Mag\\ 
      (sec)  & (sec) &  &  \\
      \hline \hline
      78373	& 1200  & OSN $B$    & $>$ 22.7 \\
      81977	& 300   & OSN $V$    & 20.96 $\pm$ 0.18 \\
      816       & 60    & INT $R$    & 17.28 $\pm$ 0.13 \\
      981       & 180   & INT $R$    & 17.31 $\pm$ 0.14 \\
      1074	& 300	& NOT $R$    & 17.45 $\pm$ 0.09 \\	
      1391      & 600   & INT $R$    & 17.44 $\pm$ 0.12 \\
      1468	& 300 	& NOT $R$    & 17.43 $\pm$ 0.08 \\	
      1853      & 180   & INT $R$    & 17.49 $\pm$ 0.12 \\
      1862	& 300	& NOT $R$    & 17.55 $\pm$ 0.09 \\	
      5363	& 120   & OSN $R$    & 16.62 $\pm$ 0.09 \\ 
      8300      & 300   & INT $R$    & 17.03 $\pm$ 0.14 \\
      18360     & 1200  & FTN $R$    & 17.90 $\pm$ 0.04 \\
      29940     & 1050  & FTN $R$    & 18.50 $\pm$ 0.02 \\
      68235     & 1200  & Asiago $R$ & 19.64 $\pm$ 0.04 \\
      75990	& 180   & OSN $R$    & 19.87 $\pm$ 0.15 \\
      80917	& 180   & OSN $R$    & 19.87 $\pm$ 0.09 \\
      82225	& 180   & OSN $R$    & 19.91 $\pm$ 0.07 \\	
      160557    & 1200  & Asiago $R$ & 20.92 $\pm$ 0.07 \\
      209760    & 960   & FTN $R$    & 21.23 $\pm$ 0.10 \\
      248617	& 1500  & OSN $R$    & 21.81 $\pm$ 0.28 \\ 
      382560    & 960   & FTN $R$    & $>$ 21.9 \\
      687323    & 120   & TNG $R$    & 23.19 $\pm$ 0.25 \\
      1121271	& 600	& NOT $R$    & 24.66 $\pm$ 0.41 \\
      2160836	& 600   & NOT $R$    & $>$ 23.6 \\
      5529	& 120   & OSN $I$    & 15.77 $\pm$ 0.12 \\
      82424	& 180   & OSN $I$    & 19.18 $\pm$ 0.15 \\
      \hline
    \end{tabular}
\end{table}

\subsection{X-ray}\label{section:obs-x}

The X-ray event data from the \emph{Swift} X-Ray Telescope (XRT; \citealt{burrows2005:SSRv120})  
were initially processed with the FTOOL, \texttt{xrtpipeline} (v0.9.9). Source and background 
spectra from 0.3 -- 10.0 keV in Windowed Timing (WT) and Photon Counting (PC) mode 
were extracted for analysis with \texttt{Xspec}, 
while the pre-reduced XRT light curve was downloaded from the on-line repository \citep{evans2007:arXiv0704}.


\section{Results}\label{section:results}

\subsection{Light curves}\label{section:results-lightcurves}

\begin{figure} 
  \centering 
  \resizebox{\hsize}{!}{\includegraphics[angle=-90]{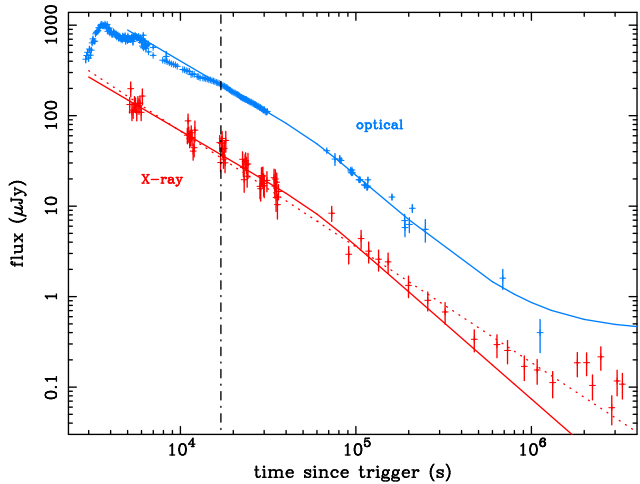} }
  \caption{Optical ($R$-band, upper blue crosses) and X-ray count rate ($\times 200$, lower red crosses) light curves of GRB\,060206. The solid lines show the smoothly broken power-law (with host correction) fit to the optical data to the right of the vertical dot-dash line, and those same parameters scaled to the X-ray. The dotted line shows a single power-law fit to the X-ray.} 
  \label{x-opt.lc} 
\end{figure}

Visual inspection of the optical light curve (Figure \ref{x-opt.lc}) clearly shows significant re-brightening at $\sim 4000$\,s and a ``bump'' at $\sim 1.7 \times 10^4$\,s, after which there is a smooth decay with a break at $\sim 5 \times 10^4$\,s \citep{wozniak2006:apj642,monfardini2006:apj648,stanek2007:apj654}.
Fitting a broken power-law to the data after the ``bump'' gives $\alpha_{1} = 1.138 \pm 0.005$,  $\alpha_{2} = 1.70 \pm 0.06$ and places the break at $t_{\mathrm{break}} = 5.9 \pm 0.5 \times 10^4$\,s ($\chi^{2}_{\nu} = 0.77$, 71 degrees of freedom, d.o.f.). 
It is plausible that the late data suffer from contamination due to the host galaxy which is estimated as $R \sim 24.6$ (Th\"{o}ne et al. in prep.) and therefore we have included this in our model.

The X-ray light curve also displays a re-brightening at $\sim 4000$\,s (e.g., \citealt{monfardini2006:apj648}) and a flattening after $\sim 10^6$\,s which has been attributed to a nearby contaminating X-ray source \citep{stanek2007:apj654}. 
We use the count rate light curve since, as we will show in Section \ref{section:results-spectra}, the X-ray data are best described by a single, unchanging spectral index, so converting to flux only adds uncertainties.
The X-ray data from $4000$ -- $10^6$\,s are well fit by a single power-law decay with $\alpha = 1.28 \pm 0.02$ ($\chi^{2}_{\nu} = 1.0$, 65 d.o.f.). 
However, we  also fit a broken power-law with $\alpha_{1} = 1.04 \pm 0.10$,  $\alpha_{2} = 1.40 \pm 0.7$ and a break time of $t_{\mathrm{break}} =  2.2^{+2.0}_{-0.8} \times 10^4$\,s ($\chi^{2}_{\nu} = 0.79$, 63 d.o.f.), giving a marginal improvement.
To test whether the X-ray is indeed consistent with the optical, we fix the temporal slopes and break time to those of the optical and fit the X-ray data. We find that these parameters well describe the X-ray data ($\chi^{2}_{\nu} = 0.94$, 66 d.o.f.; Figure \ref{x-opt.lc}). 
The results of our temporal fits are summarised in Table \ref{indices}.

\subsection{Spectral analysis}\label{section:results-spectra}

The XRT spectra were fit with an absorbed power-law and in both the WT and PC mode data, a significant amount of absorption over the Galactic value was required.  This excess extinction may be explained by host extinction in the rest frame of the burst. For the WT mode data (i.e., pre-break), 
a spectral index of $\beta_{\mathrm{X}} = 1.26 \pm 0.06$ was found ($\chi^{2}_{\nu} = 1.10$, 94 d.o.f.) while the PC mode data (i.e., post-break) was found to have
a spectral index of $\beta_{\mathrm{X}} = 0.92 \pm 0.09$ ($\chi^{2}_{\nu} = 0.93$, 59 d.o.f.).

\begin{table}	
  \centering 	
    \caption{The temporal decay indices in X-ray and optical for a single power-law, $\alpha$, and a smoothly broken power-law, $\alpha_{1}$ \& $\alpha_{2}$, with a break time, $t_{\mathrm{break}}$. Also the spectral indices for X-ray, optical and combined X-ray/optical fits (Section \ref{section:results}).} 	
    \label{indices} 	
    \begin{tabular}{l l l l} 
                                           & X-ray                   & optical    & combined \\
      \hline \hline
      $\alpha$                             & 1.28 $\pm$ 0.02         &  --                & -- \\
      $\alpha_{1}$                         & 1.04 $\pm$ 0.10         & 1.138 $\pm$ 0.005  & -- \\
      $\alpha_{2}$                         & 1.40 $\pm$ 0.07         & 1.70  $\pm$ 0.06   & -- \\
      $t_{\mathrm{break}}\times 10^4$\,s   & 2.2 $^{+2.0}_{-0.8}$    & 4.9   $\pm$ 0.5    & --  \\
      $\beta_{\mathrm{pre-break}}$         & 1.26 $\pm$ 0.06 & 0.84 $\pm$ 0.05     & 0.93 $\pm$ 0.01 \\
      $\beta_{\mathrm{post-break}}$        & 0.92 $\pm$ 0.09 & 1.4 $\pm$ 0.6       & 1.00 $\pm$ 0.06  \\
      \hline
    \end{tabular}
\end{table}

Two optical spectral indices are found by fitting the optical spectral energy distributions (SEDs) at $\sim 1.0 \times 10^4$\,s and $\sim 8.2 \times 10^4$\,s (i.e., pre- and post-break). For the pre-break analysis we use the near-infrared data ($JHK_S$) of \cite{alatalo2006:GCN4702} and the shifted $R$ band data, at that time, of \cite{stanek2007:apj654}.
For the post-break SED we use our $V$, $R$ and $I$ band data. All data were converted to fluxes and corrected for Galactic extinction of $E_{(B-V)} = 0.013$ \citep{schlege1998:ApJ500} and line blanketing due to the Lyman forest. We find optical spectral indices of $\beta_{\mathrm{opt}} = 0.84 \pm 0.05$ and $\beta_{\mathrm{opt}} = 1.4 \pm 0.6$ for pre- and post-break, respectively. 

To constrain these values further we use the simultaneous X-ray and optical fitting detailed in \cite{starling2007:ApJ661}. In this method, the optical to X-ray SED is fit in count-space, 
incorporating the measured metallicity \citep{fynbo2006:aa451} and including the effect host galaxy extinction.
The above optical data points were augmented by X-ray data at the given times: the pre-break SED by one orbit of XRT data and the post-break SED by $\sim 3 \times10^4$ seconds of data.
From this we find that both epochs are well described by a single spectral power-law with $\beta = 0.93 \pm 0.01$ and $\beta = 1.00 \pm 0.06$, respectively, in agreement with each other and with our previous values of $\beta_{\mathrm{opt}}$ and $\beta_{\mathrm{X}}$ but inconsistent with the interpretation of a possible spectral change in the optical between the two epochs. These results are shown in Table \ref{indices} and agree, within errors, with those of \cite{monfardini2006:apj648}.


\section{Discussion}\label{section:discussion}

We have shown that the well sampled X-ray afterglow can be described by a single power-law decay, though a broken power-law, which gives temporal indices and a break time similar to those in the optical, is as good a fit.
While it is difficult to accommodate the single power-law decay in the framework of the blast wave model, 
an achromatic broken power-law decay can be interpreted in terms of a jet break or an energy injection break, which we will now discuss in the context of the blast wave model (for a review and mathematical relations see, e.g., \citealt{zhang2004:IJMPA19}).

The spectral indices of the optical to X-ray spectrum are constant before and after the optical break, i.e., at $\sim 2.9$ and $\sim 23$\,hours with the break at $\sim 16$\,hours ($\sim 3$\,hours in the rest frame) after the burst. This indicates that the temporal break is not caused by the passage of a break frequency through the optical regime in the broadband spectrum. 
The conclusion one can draw from this is that the break is caused by a change in the dynamics of the jet, e.g., the cessation of the energy injection phase or the beginning of the jet-spreading phase (the jet break interpretation). 
Assuming that the optical and X-ray emission is caused by the same mechanism, the X-ray light curve is expected to show a break at the same time as the optical.

We note that \cite{monfardini2006:apj648} ascribe the dynamical change of the blast wave to a change in the circumburst density profile, the blast wave breaking out of a homogeneous medium into a stellar wind like environment. This model agrees with the observed spectral and temporal slopes but is not expected from the immediate environment models of GRB progenitors which predicts a transition from a wind like to a homogeneous medium, and not the converse (e.g., \citealt{wijers2001:grba.conf306,ramirezruiz2005:apj631}). 
In the following we explore the two possible explanations we propose for the SEDs and light curves of the afterglow of this burst, a jet break and an energy injection break.

\subsection{Jet break versus energy injection}

From the SED spectral indices the power-law index of the electron energy distribution, $p$, can be determined. 
For both possible explanations the interpretation of the SEDs is the same, in that
the single power-law SED from optical to X-rays is either in between the peak frequency, $\nu_{\rm{m}}$ and the cooling frequency, $\nu_{\rm{c}}$, or above both frequencies. 
In the first case $p = 3.00\pm 0.12$, while in the latter case $p = 2.00\pm 0.12$, using the spectral slopes from the optical to X-ray fit in count-space at $23$\,hours after the burst.

In the jet break interpretation of the achromatic break, the blast wave is moving ultra-relativistically, but decelerating, before the break.
When the Lorentz factor of the blast wave drops below the inverse half opening angle of the jet, the observer starts to see the whole jet and the jet begins to spread sideways, giving rise to the so-called jet break. 
If both the X-ray and optical regimes are situated between $\nu_{\rm{m}}$ and $\nu_{\rm{c}}$, the temporal slope before the break, given the value of $p$ derived from the SED, is $\alpha = 3(p-1)/4 = 1.50\pm 0.09$ or $\alpha = (3p-1)/4 = 2.00\pm 0.09$, for a homogeneous or a stellar wind environment, respectively. The post-break slope would then be $\alpha = p = 3.00\pm 0.12$. 
All these slopes are too steep compared to the observed temporal slopes. 
If, however, both observing regimes are above $\nu_{\rm{m}}$ and $\nu_{\rm{c}}$, the pre-break slope is $\alpha = (3p-2)/4 = 1.00\pm 0.09$, while the post-break slope is $\alpha = p = 2.00\pm 0.12$. 
The pre-break slope in this case is consistent with the observed slopes. The observed post-break slopes are slightly shallower than expected, but they are consistent within $3\sigma$, though further steepening to an asymptotic value of $\alpha = p$ cannot be ruled out. To conclude, in the jet break interpretation we find that $p = 2.00\pm 0.12$ and $\nu_{\rm{m,c}}<\nu_{\rm{opt, X}}$, but we cannot say anything about the structure of the circumburst medium, i.e., homogeneous or wind, since that requires that the observing frequencies are below $\nu_{\rm{c}}$.

If the achromatic break is interpreted as the cessation of an extended energy injection phase, the post-break slopes are given by the expressions for an ultra-relativistic blast wave. 
In this case, if both observing frequencies are situated in between $\nu_{\rm{m}}$ and $\nu_{\rm{c}}$, 
the temporal slopes after the break are $\alpha = 3(p-1)/4 = 1.50\pm 0.09$ (homogeneous medium) or $\alpha = (3p-1)/4 = 2.00\pm 0.09$ (stellar wind). 
If both observing frequencies are situated above the spectral break frequencies, the temporal slope is $\alpha = (3p-2)/4 = 1.00\pm 0.09$, regardless of the circumburst medium structure. 
Comparing these numbers with the observed post-break slopes, the observations are best fit when 
$\nu_{\rm{m}}<\nu_{\rm{opt, X}}<\nu_{\rm{c}}$ and hence $p = 3.00\pm 0.12$, and the ambient medium is homogeneous. 
Assuming that the energy injection can be described as $E\propto t^{q}$, the flattening of the light curves before the break is given by $\Delta\alpha = (p+3)/4 \times q \simeq 1.5\times q$, which gives $q\sim 0.3$ from the observed average flattening of $\Delta\alpha\sim 0.4$.

\subsection{Energetics}

In general, the jet break time is related to the half opening angle of the jet, from which the isotropic equivalent energy can be converted into the collimation corrected energy. 
If we interpret the achromatic break at $\sim 16$\,hours as a jet break, the half opening angle of the jet is found to be 
$\theta_{\rm{0}} = 0.075\times (E_{52}/n_{0})^{-1/8}\sim 4^{\circ}$ or 
$\theta_{\rm{0}} = 0.11\times (E_{52}/A_{\ast})^{-1/4}\sim 7^{\circ}$, 
for a homogeneous medium or a stellar wind environment, respectively \citep{panaitescu2002:ApJ571}. 
If we adopt the energy injection interpretation, the observations indicate that there has not been a jet break up to $10$\,days after the burst, which results in a lower limit on the jet half opening angle of 
$\theta_{\rm{0}} > 0.22\times (E_{52}/n_{0})^{-1/8}\sim 13^{\circ}$. 
In all these expressions for the opening angle, $E_{52}$ is the isotropic equivalent blast wave energy in units of $10^{52}$ ergs; $n_{0}$ is the homogeneous circumburst medium density in cm$^{-3}$; and $A_{\ast}=\dot{M}/(4\pi v_{w}^{2})$, with $\dot{M}$ the mass-loss rate in $10^{-5}$ M$_{\odot}$ per year and $v_{w}$ the stellar wind velocity in $10^3$\,km\,s$^{-1}$. 
These typical values for the energy and density \citep{panaitescu2002:ApJ571} are in agreement with the constraints on the values for $\nu_{\rm{m}}$ and $\nu_{\rm{c}}$ compared to the observing frequencies. Also the fractional energies of radiating electrons and magnetic field, $\varepsilon_{\rm{e}}$ and $\varepsilon_{\rm{B}}$ respectively, have typical values of $\sim 0.1$, although in the energy injection interpretation $\varepsilon_{\rm{B}}\sim 10^{-3.5}$, which has been found for other bursts. 
With these opening angles we can convert the isotropic equivalent gamma-ray energy of $6\times 10^{52}$\,ergs \citep{palmer2006:GCN4697} into collimation corrected energies of $2-4\times 10^{50}$\,erg for the jet break interpretation and $>10^{51}$\,ergs for the energy injection interpretation, consistent with the energy distribution of other bursts \citep{frail2001:ApJ562}.

\subsection{Implications}

Many previously studied jet breaks do not display sharp changes in the temporal decay index, but a shallow roll-over from asymptotic values which is described by a smoothly broken power-law. 
The prototypical example of such a break is GRB\,990510 for which well sampled $B$, $V$, $R$ and $I$ band light curves display an achromatic break (e.g., \citealt{stanek1999:ApJ522}). 
This is accepted as a jet break even though the X-ray light curve as measured by \emph{BeppoSAX} \citep{kuulkers2000:ApJ538} is satisfactorily described by a single power-law. 
A break at X-ray frequencies at the same time as the optical break is however, not ruled out and the temporal slopes before and after that break are similar in the optical and X-rays. 
In the analysis of GRB\,060206 we are seeing the same phenomenon: the optical light curve displays a break, while the X-ray is satisfactorily described by a single power-law fit, though a broken power-law is not ruled out.
However, an X-ray break is necessary to explain the afterglow when interpreting it in the context of the standard blast wave model. 
A similar issue has been addressed in SED fits by \cite{starling2007:arXiv0704.3718}, where adding a cooling break to some SEDs gives only a marginal improvement according to the statistical F-test, but is necessitated by considerations of the physical model.
This has significant implications for the analysis of the myriad of X-ray light curves that the \emph{Swift} satellite has afforded us. 
For those X-ray light curves extending up to $\sim 1$\,day or longer, for which we do not have well sampled optical light curves, caution is required when making claims about the absence of breaks in isolation, without considering physical interpretations.
This is particularly important when performing statistical analyses on a large sample of temporal and spectral slopes, for making collimation corrected energy estimates, and for using GRBs as standard candles.


\section{Conclusion}\label{section:conclusion}

We identify a possible achromatic break in the X-ray and optical light curves of
GRB\,060206 at $\sim 16$\,hours, which is most successfully explained by a
change in the dynamics of the jet: either as a jet break or a break due to the
cessation of energy injection.  Neither is favoured as both are consistent
with the blast wave model and the distribution of collimation corrected
energies. The presence of a weak constant source near the afterglow in
both X-rays and optical precludes, in this case, an examination of the
light curves later than $\sim 10^6$\,s.
GRB\,060206 was, up to now, assumed to have a chromatic break (i.e., a break only in
the optical) since the X-ray data alone does not require a break.
However, examining all X-ray and optical data until late times, we find
that the optical and X-ray light curves are consistent with having the same
break time and pre- and post-break temporal slopes. There is also no evidence
of chromaticity from a comparison of pre- and post-break SEDs that encompass
optical and X-ray data.

We should therefore be cautious in ruling out
breaks as being achromatic from comparing the nominal fitted slopes.
This issue is important for determining true GRB energies, but also
has a strong bearing on recent attempts to use GRBs for determining 
the geometry of the distant Universe.
That said, there does seem to be a tendency, if not yet strongly significant,
for the X-ray light curves to have less pronounced breaks. Both 060206 and
990510, the achromatic break `poster child', are examples of this. It would
therefore be worthwhile to extend the sample of \emph{Swift} bursts that have 
well sampled late-time optical light curves, which would be helped by finding
more afterglows in the anti-Sun direction. 
Also, more detailed theoretical models of jet breaks (likely involving numerical 
simulations of the jet dynamics) should be preformed to clarify whether jet 
breaks could vary somewhat between wavebands.


\section*{Acknowledgements}
We thank the referee for constructive comments.
PAC, RAMJW and KW gratefully acknowledge support of NWO under grant 639.043.302.
MRB, ER and RLCS gratefully acknowledge support from PPARC.
DM acknowledges the Instrument Center for Danish Astrophysics. 
ASJ acknowledges Wihuri foundation, Finland.
Partially supported by Spanish research programmes ESP2002-04124-C03-01 \& AYA2004-01515.
NOT operated by Denmark, Finland, Iceland, Norway, \& Sweden, in Observatorio del Roque de los Muchachos of the Instituto de Astrofisica de Canarias, Spain.D
The Dark Cosmology Centre is funded by the Danish National Research Foundation.
We acknowledge benefits from collaboration within the EU FP5 Research Training Network ``Gamma-Ray Bursts: An Enigma and a Tool" (HPRN-CT-2002-00294). 
This work made use of data supplied by the UK Swift Science Data Centre at the University of Leicester and the High Energy Astrophysics Science Archive Research Center Online Service, provided by the NASA/GSFC.

\label{lastpage}


\begin{thebibliography}{}

\bibitem[\protect\citeauthoryear{{Alatalo}, {Perley}, \& {Bloom}}{{Alatalo}
  et~al.}{2006}]{alatalo2006:GCN4702}
{Alatalo}, K., {Perley}, D.,  \& {Bloom}, J.~S. 2006, GRB Coordinates Network,
  4702

\bibitem[\protect\citeauthoryear{{Burrows} et~al.}{{Burrows}
  et~al.}{2005}]{burrows2005:SSRv120}
{Burrows}, D.~N., et~al. 2005, Space Science Reviews, 120, 165

\bibitem[\protect\citeauthoryear{{Burrows} \& {Racusin}}{{Burrows} \&
  {Racusin}}{2007}]{burrows2007:astro.ph2633}
{Burrows}, D.~N.,  \& {Racusin}, J. 2007, Il Nuovo Cimento C, in press (ArXiv
  Astrophysics e-prints 0702633)

\bibitem[\protect\citeauthoryear{{Curran} et~al.}{{Curran}
  et~al.}{2007}]{curran2006:astro.ph.10067}
{Curran}, P.~A., {Kann}, D.~A., {Ferrero}, P., {Rol}, E.,  \& {Wijers},
  R.~A.~M.~J. 2007, Il Nuovo Cimento C, in press (ArXiv Astrophysics e-prints
  0610067)

\bibitem[\protect\citeauthoryear{{Evans} et~al.}{{Evans}
  et~al.}{2007}]{evans2007:arXiv0704}
{Evans}, P.~A., et~al. 2007, A\&A, accepted (ArXiv e-prints 0704.0128)

\bibitem[\protect\citeauthoryear{{Frail} et~al.}{{Frail}
  et~al.}{2001}]{frail2001:ApJ562}
{Frail}, D.~A., et~al. 2001, {ApJ}, 562, L55

\bibitem[\protect\citeauthoryear{{Fynbo} et~al.}{{Fynbo}
  et~al.}{2006}]{fynbo2006:aa451}
{Fynbo}, J.~P.~U., et~al. 2006, {A\&A}, 451, L47

\bibitem[\protect\citeauthoryear{{Gehrels} et~al.}{{Gehrels}
  et~al.}{2004}]{gehrels2004:ApJ611}
{Gehrels}, N., et~al. 2004, {ApJ}, 611, 1005

\bibitem[\protect\citeauthoryear{{Kuulkers} et~al.}{{Kuulkers}
  et~al.}{2000}]{kuulkers2000:ApJ538}
{Kuulkers}, E., et~al. 2000, {ApJ}, 538, 638

\bibitem[\protect\citeauthoryear{{Landolt}}{{Landolt}}{1992}]{landolt1992:AJ10%
4}
{Landolt}, A.~U. 1992, {AJ}, 104, 340

\bibitem[\protect\citeauthoryear{{Madau}}{{Madau}}{1995}]{madau1995:ApJ441}
{Madau}, P. 1995, {ApJ}, 441, 18

\bibitem[\protect\citeauthoryear{{M\'esz\'aros}, {Rees}, \&
  {Wijers}}{{M\'esz\'aros} et~al.}{1998}]{meszaros1998:apj499}
{M\'esz\'aros}, P., {Rees}, M.~J.,  \& {Wijers}, R.~A.~M.~J. 1998, {ApJ}, 499,
  301

\bibitem[\protect\citeauthoryear{{Monfardini} et~al.}{{Monfardini}
  et~al.}{2006}]{monfardini2006:apj648}
{Monfardini}, A., et~al. 2006, {ApJ}, 648, 1125

\bibitem[\protect\citeauthoryear{{Nousek} et~al.}{{Nousek}
  et~al.}{2006}]{nousek2006:ApJ642}
{Nousek}, J.~A., et~al. 2006, {ApJ}, 642, 389

\bibitem[\protect\citeauthoryear{{Palmer} et~al.}{{Palmer}
  et~al.}{2006}]{palmer2006:GCN4697}
{Palmer}, D., et~al. 2006, GRB Coordinates Network, 4697

\bibitem[\protect\citeauthoryear{{Panaitescu} \& {Kumar}}{{Panaitescu} \&
  {Kumar}}{2002}]{panaitescu2002:ApJ571}
{Panaitescu}, A.,  \& {Kumar}, P. 2002, {ApJ}, 571, 779

\bibitem[\protect\citeauthoryear{{Panaitescu} et~al.}{{Panaitescu}
  et~al.}{2006}]{panaitescu2006:MNRAS369}
{Panaitescu}, A., {M{\'e}sz{\'a}ros}, P., {Burrows}, D., {Nousek}, J.,
  {Gehrels}, N., {O'Brien}, P.,  \& {Willingale}, R. 2006, {MNRAS}, 369, 2059

\bibitem[\protect\citeauthoryear{{Ramirez-Ruiz} et~al.}{{Ramirez-Ruiz}
  et~al.}{2005}]{ramirezruiz2005:apj631}
{Ramirez-Ruiz}, E., {Garc{\'{\i}}a-Segura}, G., {Salmonson}, J.~D.,  \&
  {P{\'e}rez-Rend{\'o}n}, B. 2005, {ApJ}, 631, 435

\bibitem[\protect\citeauthoryear{{Rees} \& {M\'esz\'aros}}{{Rees} \&
  {M\'esz\'aros}}{1992}]{rees1992:mnras258}
{Rees}, M.~J.,  \& {M\'esz\'aros}, P. 1992, {MNRAS}, 258, 41P

\bibitem[\protect\citeauthoryear{{Rhoads}}{{Rhoads}}{1997}]{rhoads1997:ApJ487}
{Rhoads}, J.~E. 1997, {ApJ}, 487, L1

\bibitem[\protect\citeauthoryear{{Rhoads}}{{Rhoads}}{1999}]{rhoads1999:ApJ525}
{Rhoads}, J.~E. 1999, {ApJ}, 525, 737

\bibitem[\protect\citeauthoryear{{Sari}, {Piran}, \& {Halpern}}{{Sari}
  et~al.}{1999}]{sari1999:ApJ519}
{Sari}, R., {Piran}, T.,  \& {Halpern}, J.~P. 1999, {ApJ}, 519, L17

\bibitem[\protect\citeauthoryear{{Schlegel}, {Finkbeiner}, \&
  {Davis}}{{Schlegel} et~al.}{1998}]{schlege1998:ApJ500}
{Schlegel}, D.~J., {Finkbeiner}, D.~P.,  \& {Davis}, M. 1998, {ApJ}, 500, 525

\bibitem[\protect\citeauthoryear{{Stanek} et~al.}{{Stanek}
  et~al.}{2007}]{stanek2007:apj654}
{Stanek}, K.~Z., et~al. 2007, {ApJ}, 654, L21

\bibitem[\protect\citeauthoryear{{Stanek} et~al.}{{Stanek}
  et~al.}{1999}]{stanek1999:ApJ522}
{Stanek}, K.~Z., {Garnavich}, P.~M., {Kaluzny}, J., {Pych}, W.,  \& {Thompson},
  I. 1999, {ApJ}, 522, L39

\bibitem[\protect\citeauthoryear{{Starling} et~al.}{{Starling}
  et~al.}{2007a}]{starling2007:arXiv0704.3718}
{Starling}, R.~L.~C., {Van der Horst}, A.~J., {Rol}, E., {Wijers}, R.~A.~M.~J.,
  {Kouveliotou}, C., {Wiersema}, K., {Curran}, P.~A.,  \& {Weltevrede}, P.
  2007a, ArXiv e-prints 0704.3718

\bibitem[\protect\citeauthoryear{{Starling} et~al.}{{Starling}
  et~al.}{2007b}]{starling2007:ApJ661}
{Starling}, R.~L.~C., {Wijers}, R.~A.~M.~J., {Wiersema}, K., {Rol}, E.,
  {Curran}, P.~A., {Kouveliotou}, C., {Van der Horst}, A.~J.,  \& {Heemskerk},
  M.~H.~M. 2007b, {ApJ}, 661, 787

\bibitem[\protect\citeauthoryear{{Wijers}}{{Wijers}}{2001}]{wijers2001:grba.co%
nf306}
{Wijers}, R.~A.~M.~J. 2001, in Gamma-ray Bursts in the Afterglow Era, ed.
  E.~{Costa}, F.~{Frontera}, \& J.~{Hjorth}, 306

\bibitem[\protect\citeauthoryear{{Wo{\'z}niak} et~al.}{{Wo{\'z}niak}
  et~al.}{2006}]{wozniak2006:apj642}
{Wo{\'z}niak}, P.~R., {Vestrand}, W.~T., {Wren}, J.~A., {White}, R.~R.,
  {Evans}, S.~M.,  \& {Casperson}, D. 2006, {ApJ}, 642, L99

\bibitem[\protect\citeauthoryear{{Zhang} \& {M{\'e}sz{\'a}ros}}{{Zhang} \&
  {M{\'e}sz{\'a}ros}}{2004}]{zhang2004:IJMPA19}
{Zhang}, B.,  \& {M{\'e}sz{\'a}ros}, P. 2004, International Journal of Modern
  Physics A, 19, 2385

\end{thebibliography}
\end{document}